\documentclass[12pt]{article}
\usepackage{amstex}
\usepackage{amssymb}
\begin{document}
\begin{center}
\LARGE
\textbf{On Bohmian trajectories in two-particle
interference devices}\\[1cm]
\large
\textbf{Louis Marchildon}\\[0.5cm]
\normalsize
D\'{e}partement de physique,
Universit\'{e} du Qu\'{e}bec,\\
Trois-Rivi\`{e}res, Qc.\ Canada G9A 5H7\\
email: marchild$\hspace{0.3em}a\hspace{-0.8em}
\bigcirc$uqtr.uquebec.ca\\
\end{center}
\medskip
\begin{abstract}
Claims have been made that, in two-particle
interference experiments involving bosons, Bohmian
trajectories may entail observable consequences 
incompatible with standard quantum mechanics.
By general arguments and by an examination of
specific instances, we show that this is not the
case.\\[1ex]
\end{abstract}
\section{Introduction}
The question whether quantum mechanics can be derived
from a deterministic theory has been raised for a
long time~\cite{einstein,neumann}, and answered
in the affirmative by Bohm~\cite{bohm}.  In Bohmian
mechanics~\cite{holland,hiley}, each particle follows a 
trajectory that obeys equations of motion much like
those of classical mechanics.  To the total force acting
on a particle, however, there is a contribution
coming from a ``quantum potential,'' which is related
to the amplitude of the Schr\"{o}dinger wave function.
Although a particle has, at any time $t$, a
well-defined position and momentum, these cannot be
known exactly.  One can only know the probability
that, at time $t$, the particle is in a given region
of space, or has a momentum in a given range.  
The position probability is obtained from the
absolute square of the wave function and, in this way,
Bohmian mechanics reproduces exactly the statistical
predictions of quantum mechanics.

Recent papers~\cite{ghosea,ghoseb,golshani} have
argued that although statistical predictions
of Bohmian and quantum mechanics may coincide,
the two will disagree as far as certain individual events
are concerned.  This, it is claimed, will occur in
two-particle interference experiments, where Bohmian 
mechanics would predict correlations in individual
events that contradict quantum mechanics.  The purpose
of this note is firstly to give a general argument
that this cannot be the case, and then to show how the
specific experiments proposed fail to establish the
intended conclusion.
\section{Two-particle interference}
\begin{figure}[ht]
\centering
\begin{picture}(260,100)(0,0)
\multiput(35,50)(15,0){13}{\line(1,0){10}}
\multiput(2,50)(0,30){2}{\line(1,0){6}}
\multiput(5,50)(0,20){2}{\line(0,1){10}}
\multiput(230,5)(0,10){10}{\line(2,-1){10}}
\thicklines
\put(30,0){\line(0,1){15}}
\put(30,25){\line(0,1){50}}
\put(30,85){\vector(0,1){15}}
\put(230,0){\line(0,1){100}}
\thinlines
\qbezier(30,80)(129,72.5)(228,65)
\qbezier(30,20)(129,30)(228,40)
\put(245,50){\vector(1,0){15}}
\put(228,65){\makebox(0,0){$\bullet$}}
\put(228,40){\makebox(0,0){$\bullet$}}
\put(5,65){\makebox(0,0){$a$}}
\put(22,80){\makebox(0,0){$A$}}
\put(22,20){\makebox(0,0){$B$}}
\put(20,95){\makebox(0,0){$x$}}
\put(254,42){\makebox(0,0){$y$}}
\end{picture}
\caption{Two-slit interferometer.  Solid lines are
schematic, and do not represent the exact shape of
Bohmian trajectories.}
\label{f1}
\end{figure}
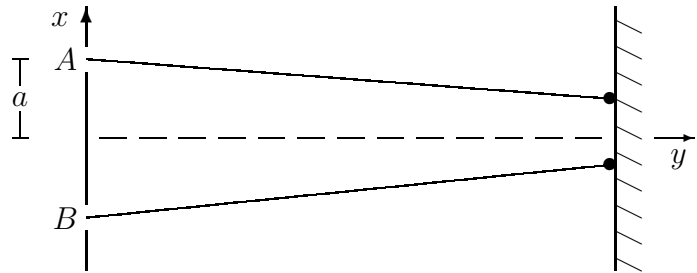

The general situation can be developed in terms of a
two-slit interferometer, as shown in Fig.~\ref{f1}.
Two identical bosons are prepared in a state such
that, at time $t = 0$, one is near the upper and the
other one near the lower slit.  The system is assumed
to be symmetric with respect to the $yz$ plane.

The wave function can be written as
\begin{equation}
\Psi (\mathbf{r}_1, \mathbf{r}_2)
= \psi_A (\mathbf{r}_1) \psi_B (\mathbf{r}_2)
+ \psi_A (\mathbf{r}_2) \psi_B (\mathbf{r}_1) .
\label{eq1}\end{equation}
As it should with bosons, it is symmetric
with respect to the interchange of both particles.
An alternative choice for the wave function is
given by
\begin{equation}
\tilde{\Psi} (\mathbf{r}_1, \mathbf{r}_2)
= [\tilde{\psi}_A (\mathbf{r}_1) + \tilde{\psi}_B (\mathbf{r}_1)]
[\tilde{\psi}_A (\mathbf{r}_2) + \tilde{\psi}_B (\mathbf{r}_2)] .
\label{eq2}\end{equation}
This allows both particles to go through
the same slit.  Note that (\ref{eq2}) is a special
case of (\ref{eq1}) if we set $\psi_B = \psi_A$ and
no longer require $\psi_A$ to be centered about a
specific slit.

In quantum mechanics, the probability that one
particle is detected in region $R_1$ and the other
particle in region $R_2$ on the screen at time $t$
is given by
\begin{equation}
P(R_1, R_2; t)
= \int _{R_1} d\mathbf{r}_1 \int _{R_2} d\mathbf{r}_2
|\Psi (\mathbf{r}_1, \mathbf{r}_2; t)| ^2 .
\label{eq3}\end{equation}
This, in general, will display interference patterns.

In Bohmian mechanics, each particle of a given pair
has a well-defined trajectory associated with the
following velocities:
\begin{align}
& \mathbf{v}_1 = \frac{\hbar}{m} \mbox{Im} 
\frac{\boldsymbol{\nabla}_1 \Psi}{\Psi}
= \frac{1}{m} \boldsymbol{\nabla} _1 S , \label{eq4} \\
& \mathbf{v}_2 = \frac{\hbar}{m} \mbox{Im} 
\frac{\boldsymbol{\nabla}_2 \Psi}{\Psi}
= \frac{1}{m} \boldsymbol{\nabla} _2 S . \label{eq5}
\end{align}
Here $S(\mathbf{r}_1, \mathbf{r}_2; t)$ is the phase of the 
total wave function, in units of $\hbar$.

The initial wave functions of the two bosons
are assumed to transform into each other 
under reflection.  That is
\begin{equation}
\psi_A (\mathbf{r}) = \psi_B (\mathbf{r} ') ,
\label{eq6}\end{equation}
where $\mathbf{r} '$ is obtained from $\mathbf{r}$ by
reflection in the plane of symmetry, specifically
$x' = -x$, $y' = y$, $z' = z$.  Since the experimental 
arrangement shares that symmetry, the wave functions
will satisfy (\ref{eq6}) at any time $t$.  The Bohmian
trajectories of a pair of bosons, however, will
not in general transform into each other under
reflection.  This comes from the fact that the initial
values of the position of each boson are unknowable
in principle, and are each statistically distributed
according to the absolute square of the wave function.

It is not difficult to show that
\begin{align}
& v_{1x} (\mathbf{r} _1, \mathbf{r} _2; t)
= - v_{1x} (\mathbf{r} _1 ', \mathbf{r} _2 '; t) , \label{eq7}\\
& v_{2x} (\mathbf{r} _1, \mathbf{r} _2; t)
= - v_{2x} (\mathbf{r} _1 ', \mathbf{r} _2 '; t) . \label{eq8}
\end{align}
This implies that, if both particles are simultaneously
on the plane of symmetry, both velocities vanish, and
neither particle should cross the plane.  If that were 
always so, Bohmian mechanics would predict that one
particle would always be detected at positive, and the
other one at negative values of $x$.  However, the
overwhelming majority of pairs are not simultaneously
on the plane of symmetry.  (\ref{eq7}) and (\ref{eq8})
therefore do not prevent them from crossing the plane.

A related kind of restriction to the motion of
particles can be derived by considering the phase
of the total wave function.  Let us write it as
$S(\mathbf{r}, \mathbf{R}; t)$, where
$\mathbf{r}$ and $\mathbf{R}$ are the relative and
center-of-mass coordinates of the two particles.  From
(\ref{eq4}) and (\ref{eq5}), it is easy to see that
\begin{equation}
\mathbf{v}_1 + \mathbf{v}_2 
= \frac{1}{m} \boldsymbol{\nabla} _1 S 
+ \frac{1}{m} \boldsymbol{\nabla} _2 S
= \frac{1}{m} \boldsymbol{\nabla} _{\mathbf{R}} S .
\end{equation}
Thus, if the phase does not depend on the center-of-mass
coordinate $X$, we have $v_{1x} + v_{2x} = 0$, so that
\begin{equation}
x_1 + x_2 = 2 \bar{x} ,
\label{eq10}\end{equation}
where $\bar{x}$ is a constant.  This means that the motion
of the particles along the $x$ axis is symmetric with
respect to the plane $x = \bar{x}$.  Suppose there is 
a limit to the value of $|\bar{x}|$.  Bohmian mechanics
then predicts that there will be no pair of particles
detected both above $|\bar{x}|$ or both below $- |\bar{x}|$.
This has been taken to imply a contradiction 
with orthodox quantum mechanics
where, it is claimed, there is a nonzero probability
of finding both particles in one of these regions.

That there cannot be a contradiction of this kind can be 
seen as follows.  Suppose for instance that Bohmian 
mechanics implies that two particles 
can never be simultaneously detected
above the plane $x = \bar{x}$.  This is equivalent to the
statement that the probability of both particles being
above $\bar{x}$ vanishes.  In Bohmian mechanics, that
probability is given by an expression like (\ref{eq3}), 
where $R_1$ and $R_2$ are the regions $x_1 > \bar{x}$ and
$x_2 > \bar{x}$, and $|\Psi (\mathbf{r} _1, 
\mathbf{r} _2 ; t)|^2$ is the proportion of pairs whose 
true values of position at $t$ are
$\mathbf{r} _1$ and $\mathbf{r} _2$.  Since the probability
is computed in quantum mechanics with the same formula and
with a $\Psi$ which, albeit differently interpreted, has
the same numerical value, it must vanish under exactly the
same conditions as the Bohmian probability.

We shall now examine some specific cases and see
how the agreement comes about.
\section{Specific cases}
Our first example is to take $\psi _A$ and $\psi _B$
to be given by plane waves.  This is the case discussed
by Ghose~\cite{ghosea,ghoseb}.  At time $t = 0$, we set
\begin{equation}
\psi _A (\mathbf{r}) = \exp \{i(k_x x + k_y y)\} ,
\end{equation}
where for simplicity $z$ has been eliminated.
From (\ref{eq6}) we get
\begin{equation}
\psi _B (\mathbf{r}) = \exp \{i(- k_x x + k_y y)\} .
\end{equation}
Assuming free propagation with time, we find that
the total wave function (\ref{eq1}) is given by
\begin{align}
\Psi (\mathbf{r}_1, \mathbf{r}_2; t)
&= \psi_A (\mathbf{r}_1; t) \psi_B (\mathbf{r}_2; t)
+ \psi_A (\mathbf{r}_2; t) \psi_B (\mathbf{r}_1; t) \notag\\
&= 2 \cos \{k_x (x_1 - x_2)\} 
\exp \left\{ i \left[ k_y (y_1 + y_2)
- \frac{\hbar}{m} (k_x^2 + k_y^2) t \right] \right\} .
\end{align}

It is easy to show that not only does
(\ref{eq10}) hold, but $x_1$ and $x_2$ are separately
constant.  However, the marginal probability 
densities of $x_1$ and $x_2$ are both uniformly 
distributed.  This means that for each pair of particles,
the initial (and final) values of $x_1$ and $x_2$ are in no
way constrained to lie on both sides of the plane
$x = 0$, or any given plane $x = \mbox{constant}$
for that matter.  The fact that both particles can be
above, or below, the plane $x = 0$ also holds if plane
waves are replaced with spherical 
waves~\cite{marchildon,ghosec}.  The agreement between
Bohmian and quantum mechanics here comes from the fact
that although for each pair there is a plane $x = \bar{x}$
about which both particles have symmetrical $x$ coordinates,
the value of $\bar{x}$ changes from pair to pair, and is
totally unknown for any given pair.

Of course, plane waves will not represent a realistic
two-slit experiment, precisely because the $x$ coordinates
have to be restricted at $t = 0$.  But the example is
instructive, and shows that with no restrictions
on the initial $x$ positions of the particles, they will
end up in all regions on the screen.

We will soon analyse a situation with waves initially
restricted to the width of the slits.  But let us first
examine a one-dimensional case of restriction, using
harmonic oscillator wave functions.  We write
\begin{align}
\psi _A (x; t) & \sim \exp \left\{ - \frac{m \omega}{2 \hbar}
(x - a \cos \omega t) ^2 \right. \notag\\
& \qquad \left. \mbox{} - \frac{i}{2} 
\left[ \omega t + \frac{m \omega}{2 \hbar}
(4 x a \sin \omega t 
- a^2 \sin 2 \omega t) \right] \right\} .
\end{align}
This represents a wave packet of width
$\sqrt{\hbar / 2 m \omega}$ whose center oscillates
between $x = a$ and $x = -a$.  
Following (\ref{eq6}), we set 
$\psi _B (x; t) = \psi _A (-x; t)$.  For 
Bose-Einstein statistics, that is, for a total wave 
function given as in (\ref{eq1}), Holland 
(\cite{holland}, p.~300 ff.) shows that 
the phase does not depend on $x_1 + x_2$, so that
(\ref{eq10}) holds.  For each pair
of oscillators, there is a plane $x = \bar{x}$
that separates the two oscillators, so
that their trajectories do not cross.

Let us assume that $\sqrt{\hbar / 2 m \omega} \ll a$,
that is, the width of the packet is much smaller than
the amplitude of oscillation.  Most values of $|\bar{x}|$
are then smaller than or on the order of
$\sqrt{\hbar / 2 m \omega}$.  We can see how the
agreement between Bohmian and quantum mechanics comes
about.  For $t$ such that $|a \cos \omega t| \gg
\sqrt{\hbar / 2 m \omega}$, members of a given Bohmian
pair of oscillators are on different sides of the plane
$x = 0$.  But then the quantum wave packets are also widely 
separated. For $t$ such that $|a \cos \omega t| \approx
\sqrt{\hbar / 2 m \omega}$, the wave packets overlap.
But then two Bohmian oscillators on different sides of
an $x = \bar{x}$ plane can be on the same side of the
$x = 0$ plane.

Let us now turn to a more realistic description
of two-slit interference.
A wave packet emerging from slit $A$ can be
modelled by a Gaussian wave function of the type
\begin{equation}
\tilde{\psi} _A (\mathbf{r})
= (2 \pi \sigma _0^2)^{-1/4} \exp \left\{ 
- \frac{(x-a)^2}{4 \sigma _0^2} 
+ i [k_x (x-a) + k_y y] \right\} ,
\end{equation}
where $\sigma _0$ corresponds to the half-width
of the slit and $a$ is shown in Fig.~\ref{f1}.
We take $\sigma _0 \ll a$.  Assuming free 
propagation with time, we have
\begin{align}
\tilde{\psi} _A (\mathbf{r}; t)
&= (2 \pi \sigma _t^2)^{-1/4} \exp \left\{ 
- \frac{[x-a-(\hbar k_x /m) t]^2}{4 \sigma _0 \sigma _t}
\right. \notag\\
& \qquad \left. \mbox{} + i \left\{k_x [x-a-(\hbar k_x /2m)t] 
+ k_y y - (\hbar k_y^2/2m) t \right\} 
\mbox{\rule[-1.5ex]{0mm}{4.5ex}} \right\} ,
\label{eq16}\end{align}
where
\begin{equation}
\sigma _t = \sigma _0 \left( 1 + \frac{i \hbar t}
{2 m \sigma _0^2} \right) .
\end{equation}
It is shown in~\cite{golshani} that, if (\ref{eq2})
represents the total wave function and 
$\tilde{\psi} _B (\mathbf{r}; t)$ is given by 
(\ref{eq6}), we have (in~\cite{golshani} $x$ and $y$
are interchanged and the term $2 i k_x$ is
in the end inadvertently omitted)
\begin{align}
v_{1x} + v_{2x} 
&= \frac{(\hbar/2 m \sigma _0^2)^2 (x_1 + x_2) t}
{1 + (\hbar t/2 m \sigma _0^2)^2}
+ \frac{\hbar}{m} \mbox{Im} \left\{ \frac{1}{\tilde{\Psi}}
\left[ \frac{a + (\hbar k_x / m) t}{\sigma _0 \sigma _t}
+ 2 i k_x \right] \right. \notag\\
& \qquad \left. \cdot \left[ \tilde{\psi} _A (\mathbf{r} _1; t)
\tilde{\psi} _A (\mathbf{r} _2; t)
- \tilde{\psi} _B (\mathbf{r}_1; t)
\tilde{\psi} _B (\mathbf{r} _2; t ) \right]
\mbox{\rule[-1.5ex]{0mm}{4.5ex}}\right\} .
\label{eq18}\end{align}

To establish a contradiction between Bohmian mechanics
and quantum mechanics, Golshani and Akhavan consider 
separately the case where both particles go through
different slits and the case where they go through
the same slit.  With the wave function (\ref{eq2}),
that distinction has meaning only in Bohmian mechanics,
since in orthodox quantum mechanics we cannot assert
that a particle has gone through a slit unless it has
been measured to do so.

That the two particles go through different slits, in
Bohmian mechanics, means the following:  At $t = 0$,
\begin{equation}
a - \sigma _0 \lesssim x_1 \lesssim a + \sigma _0
\quad \mbox{and} \quad
-a - \sigma _0 \lesssim x_2 \lesssim -a + \sigma _0 ,
\end{equation}
or vice versa.  Contrary to the claim made 
in~\cite{golshani}, this does not imply that 
$\psi _A (\mathbf{r} _1) = \psi _B (\mathbf{r} _2)$ or that
$\psi _A (\mathbf{r} _2) = \psi _B (\mathbf{r} _1)$.

The case where both particles go through different
slits is best discussed in terms of wave function
(\ref{eq1}), where Bohmian mechanics definitely
says so and quantum mechanics predicts with certainty
(neglecting exponential tails of the wave function)
that a measurement made at $t = 0$ finds both particles
at different slits.  There (\ref{eq18}) holds with the
last term absent.  It is readily integrated 
as~\cite{golshani}
\begin{equation}
x_1 + x_2 = 2 \bar{x} 
\sqrt{1 + (\hbar t/ 2 m \sigma _0^2)^2} ,
\label{eq20}\end{equation}
where $2 \bar{x} = x_1(0) + x_2(0)$.  Note that
$- \sigma _0 \lesssim \bar{x} \lesssim \sigma _0$,
corresponding to the spread in values of $x_1(0)$ and
$x_2(0)$.  Does the constraint (\ref{eq20}) imply
observational consequences that contradict
quantum mechanics?

That there are none can best be seen by examining
the following limiting cases, where $t_f$
is the time of arrival at the screen: 
(i) $\hbar t_f/ 2 m \sigma _0^2 \ll 1$, and
(ii) $\hbar t_f/ 2 m \sigma _0^2 \gg 1$.  For
simplicity, we assume that $k_x = 0$.
In (i), we have $x_1(t_f) + x_2(t_f) \approx 2 \bar{x}$.
This suggests that $x_1(t_f)$ and $x_2(t_f)$ remain widely
separated.  But this is also what quantum mechanics
predicts, since the wave functions $\psi _A$ and $\psi _B$
do not really spread beyond $\sigma _0$.  In (ii), on
the other hand, equation (\ref{eq16}) shows that
the spread of the wave functions at $t_f$ 
is on the order of $\hbar t_f/ m \sigma _0$.  There is
overlap if the spread is on the order of $a$, 
and quantum mechanics no longer predicts
that both particles are on different sides of the $x = 0$
plane.  But then so does Bohmian mechanics, since
$x_1(t_f) + x_2(t_f) \approx \bar{x} \hbar t_f/ m 
\sigma _0^2 \approx a$.
\section{Conclusion}
We have shown that there is no reason to expect
discrepancies between Bohmian and quantum mechanics
in the context of two-particle interference devices.
Our general argument was illustrated with analysis
of specific cases.  Additional insight could be
obtained by detailed numerical calculations of
Bohmian trajectories associated with two-particle
two-slit experiments.

It is a pleasure to thank Gianluca Introzzi for help
in clarifying some of the issues discussed here.
\end{document}